# Origin of the Pseudogap in High-Temperature Cuprate Superconductors

Jamil Tahir-Kheli and William A. Goddard III

*Materials and Process Simulation Center (MC 139-74)*
*California Institute of Technology, Pasadena CA 91125*

jamil@wag.caltech.edu, wag@wag.caltech.edu

ABSTRACT Cuprate high-temperature superconductors exhibit a pseudogap in the normal state that decreases monotonically with increasing hole doping and closes at x ≈ 0.19 holes per planar $CuO_2$ while the superconducting doping range is 0.05 < x < 0.27 with optimal $T_c$ at x ≈ 0.16. Using ab initio quantum calculations at the level that leads to accurate band gaps, we found that four-Cu-site plaquettes are created in the vicinity of dopants. At x ≈ 0.05 the plaquettes percolate, so that the Cu $d_{x2y2}$/O pσ orbitals inside the plaquettes now form a band of states along the percolating swath. This leads to metallic conductivity and below $T_c$ to superconductivity. Plaquettes disconnected from the percolating swath are found to have degenerate states at the Fermi level that split and lead to the pseudogap. The pseudogap can be calculated by simply counting the spatial distribution of isolated plaquettes, leading to an excellent fit to experiment. This provides strong evidence in favor of inhomogeneous plaquettes in cuprates.

The origin of the pseudogap (PG)[1] is a major unsolved problem in cuprate superconductivity. It is believed by many to hold the key to understanding the anomalous normal state and superconductivity in these materials. The PG is a reduced density of states in the vicinity of the Fermi level in the *normal* state of cuprate high-temperature superconductors. The effect of the PG is seen in all measurements of the cuprates. The PG is largest for doping x ~ 0.05, where x is the number of holes per planar $CuO_2$. The PG decreases approximately linearly as x increases and its magnitude is ~ 80 meV for x ≈ 0.05 where $T_c$ and the superconducting gap are almost zero. Bulk measurements find that the PG goes to zero at x ≈ 0.19,[2-4] whereas the maximum doping for superconductivity is at x ≈ 0.27. In addition, STM measurements[5-8] find a spatially varying PG with an average value that decreases with doping. Since the PG appears in the normal state above the superconducting $T_c$ and vanishes at a doping value inside the superconducting phase, understanding its origin is crucial to determining the mechanism of superconductivity and the anomalous normal state properties of cuprates.

There are two classes of current rationalizations for the PG. The first is that it is a precursor Cooper pairing superconducting gap in the normal state[9-11] arising from superconducting fluctuations. In this scenario, the PG is approximately particle-hole symmetric around the Fermi level. Recent measurements[12-15] have found significant asymmetry of the PG with respect to the Fermi level and a nonzero PG at temperatures above the observed superconducting fluctuations. The second class claims the PG is a competing order to superconductivity (time-reversal, nematic, density wave, or magnetic symmetry breaking[7,8,16-34]). Our explanation here falls into this class. With no adjustable parameters, our model finds that the PG closes at x ≈ 0.19. With exactly one parameter, we predict the PG as a function of doping. This parameter is a multiplicative energy constant that sets the overall energy scale of the PG, but does not change the shape of the PG curve.

In this report, we build upon our prior work of a Plaquette model of cuprates[35-39] to explain the nature of the PG and derive its doping dependence. We first describe the plaquette model and then the detailed derivation of the PG evolution with doping.

The essential features of the plaquette model[35-39] are shown in detail in Figures 1, 2, and Supporting Information figure S1 and are summarized here:

1.) the highest level ab initio density functional calculations on undoped and explicitly doped cuprates find the hole arising from the doping to be out of the $CuO_2$ plane with predominantly apical O $p_z$ character that delocalizes over a four-Cu-site square surrounding the dopant. This four-Cu square is called a plaquette and the Cu atoms inside plaquettes are called doped Cu sites.

2.) For x ≈ 0.05 – 0.06 (experiment is ≈ 0.05), a percolating pathway or swath comprised of adjacent plaquettes is formed.

3.) This leads to a Cu $d_{x2y2}$/O pσ metallic band of states delocalized inside this swath of percolating plaquettes, which are well described with standard band methods.

4.) The undoped Cu sites remain localized $d^9$ spins with the undoped antiferromagnetic coupling between neighboring $d^9$ spins, $J_{dd}$ = 0.13 eV.

5.) Superconducting d-wave Cooper pairing of metallic Cu $x2y2$/O pσ electrons occurs by coupling to the antiferromagnetic undoped $d^9$ spins at the surface of the percolating metallic swath. Thus the onset of metallic percolation is also the onset of superconductivity and the superconducting gap is a surface to metallic volume effect.

6.) At x ≈ 0.271 (experiment ≈ 0.27), undoped $d^9$ spins no longer have an undoped $d^9$ neighbor leading to the vanishing of the superconducting pairing and no superconductivity for x > 0.271.

7.) The pseudogap arises from the splitting of degenerate Px and Py states shown in Figure 3 at the Fermi level that reside inside isolated plaquettes (not part of the percolating plaquette swath).

8.) Simple site-counting calculations explain a wide range of properties of the cuprates, such as the neutron spin resonance ("the 41 meV peak") and the universal room-temperature thermopower with no adjustable parameters.[37-39] It also provides qualitative explanations for the linear resistivity and the temperature dependent Hall effect.[37]

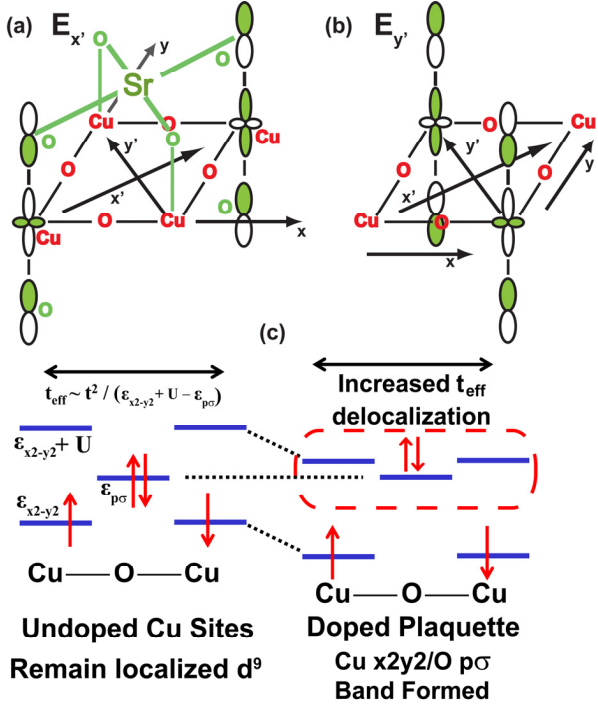

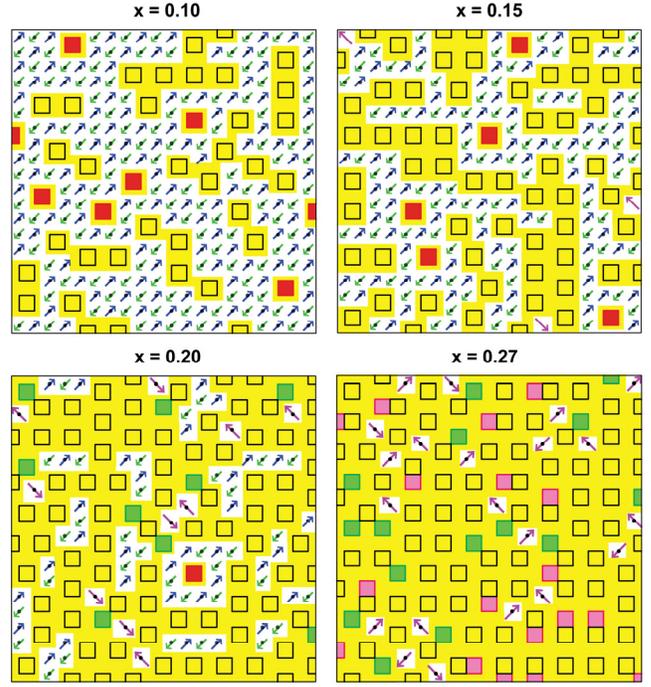

**Figure 1.** Four-site-plaquette formation in the vicinity of a Sr dopant in cuprate $La_{2-x}Sr_xCuO_4$ as found by high level ab initio calculations.[36] The two degenerate E states that are localized out-of-the $CuO_2$ plane are shown in (a) and (b). The additional hole induced by substituting Sr for La does not go into the planar Cu x2y2/O pσ orbitals as is usually assumed. Instead, it is comprised of predominantly apical O $p_z$, and Cu $d_{z^2}$. Planar atoms are red and out-of-plane orbitals and atoms are green. The planar Cu sites in (a) and (b) in the vicinity of the dopant are called doped sites. Planar Cu sites that are not in the vicinity of a dopant are called undoped sites. In (c), the energy level diagram shows how an out-of-plane hole above the planar Cu sites leads to delocalization of the doped $d^9$ Cu spins and the formation of delocalized Cu x2y2/O pσ states inside the plaquette. The left figure shows the energy levels that occur at undoped Cu sites where there are no holes in out-of-the-plane orbitals. In this case, the difference in energy between the doubly occupied Cu site and the planar O orbital energy, $\varepsilon_{x2y2} + U - \varepsilon_p$, is large, leading to localization of spins on the Cu. In the right figure, the Cu orbital energy is reduced due to the missing electron in the apical O sites directly above the Cu atoms leading to the neighboring doubly-occupied O pσ electrons delocalizing onto the Cu sites. When two plaquettes are neighbors, the delocalization occurs over all eight Cu sites. When the doping is large enough that the plaquettes percolate in 3D through the crystal, a "metallic" band is formed in the percolating swath and current can flow from one end of the crystal to the other. Further details are in the Supporting Information.

In figure 2, the number of isolated (red) plaquettes decreases as doping increases and goes rapidly towards zero at x ≈ 0.19. This can be seen in the figure where there is only one isolated plaquette in the x = 0.20 lattice and none for x = 0.27. The isolated plaquettes create the pseudogap. D-wave superconducting pairing inside the metallic swath occurs at the surface due to interaction of Cu x2y2/O pσ Cooper pairs with the neighboring antiferromagnetic undoped Cu $d^9$ spins.[37-39] The magnitude of the superconducting pairing is therefore proportional to the metallic swath's surface to volume ratio. The local pairing occurs whenever two neighboring $d^9$ antiferromagnetically coupled spins (green and blue arrows) are at the surface. The isolated Cu $d^9$ spins (purple) do not contribute to superconducting pairing. At x ≈ 0.271, there are only isolated Cu $d^9$ spins leading to the end of the superconducting phase (experiment is ≈ 0.27). We distribute plaquettes with the constraint of no overlaps (do not share a common Cu), but otherwise the distribution is completely random (see Supporting Information and S1 for details). Above ≈ 0.187 doping, it is impossible to add dopants such that the plaquettes do not overlap. The green plaquettes dope three $d^9$ spins. At ≈ 0.226 doping, this is no longer possible and the pink plaquettes dope two $d^9$ sites.

**Figure 2.** Illustration of the electronic structure in a single $CuO_2$ plane for doping values, x = 0.10, 0.15, 0.20, and 0.27 on a 20 x 20 lattice. Four kinds of electrons are shown. The green and blue arrows are undoped Cu $d^9$ spins that are antiferromagnetically coupled with the undoped superexchange coupling, $J_{dd}$ ≈ 0.13 eV. The purple arrows are undoped Cu $d^9$ spins that have no neighboring $d^9$ spin. Each black, green, pink, and red square represents a plaquette centered at a dopant and comprised of out-of-plane orbitals (apical O $p_z$ and Cu $dz^2$). The corners of the squares are planar Cu sites. The orbitals and O atoms are not shown. The black squares are plaquettes that are adjacent to another plaquette. A Cu x2y2/O pσ metallic band forms inside the percolating region of the plaquettes and is shown in yellow. For x = 0.10 and 0.15, the doping is below the 2D percolation threshold of ≈ 0.151, and the percolation occurs in 3D between $CuO_2$ layers (not shown). For x = 0.20 and 0.27, 2D percolation of the plaquettes can be directly seen. The red squares are isolated plaquettes with no plaquette neighbors. Delocalization of the Cu x2y2/O pσ electrons also occurs inside the red plaquettes as shown by the yellow surrounding the red square, but they are not connected to the percolating "metallic" swath.

Hence, cuprates are intrinsically inhomogeneous with four types of electrons: 1.) the undoped Cu $d^9$ electrons that are antiferromagnetically coupled to each other, 2.) the delocalized Cu x2y2/O pσ band electrons formed on doped sites inside the percolating plaquette region, 3.) the delocalized Cu x2y2/O pσ electrons inside isolated plaquettes, and 4.) the out-of-plane holes that only delocalize over the 4-site region in the vicinity of the dopant.

In Figure 2, the number of isolated plaquettes (red squares) has almost vanished by x = 0.20. This sharp decrease in isolated plaquettes is the origin of the vanishing PG at x ≈ 0.19, as we show below.

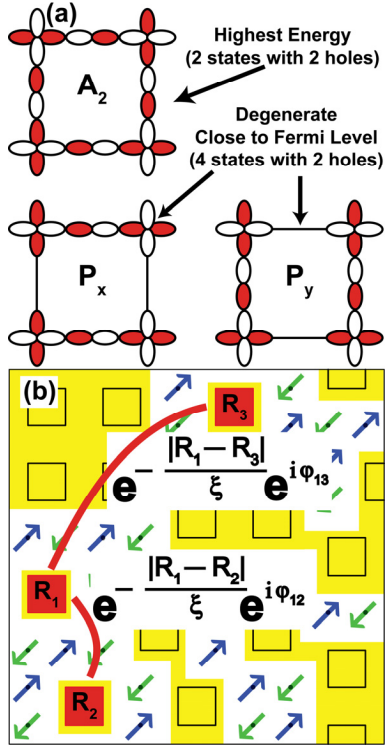

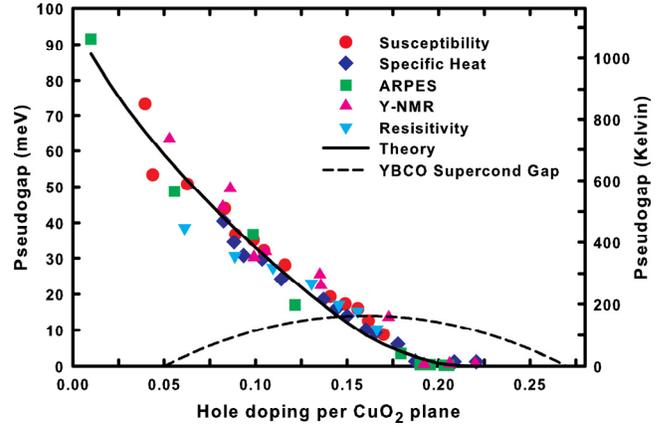

**Figure 3.** (a) The degenerate $P_x$ and $P_y$ states in the isolated plaquettes at the Fermi level. There are eight O electrons (two per O pσ) and four Cu x2y2 electrons leading to a total of twelve electrons for sixteen states. In (a), the highest unoccupied state (two holes or zero electrons) and the doubly occupied (two electrons or two holes) degenerate $P_x$ and $P_y$ plaquette states are shown (Figure S2 shows all the orbitals). (b) A 10x10 lattice subset of x = 0.15 from Figure 2 showing the exponential distance dependence of the matrix element coupling $P_x$ and $P_y$ inside the isolated plaquette at $R_1$ arising from interaction with the isolated plaquettes at $R_2$ and $R_3$. The spin-spin correlation length, ξ, is the antiferromagnetic decay length and the phases, $\varphi_{12}$ and $\varphi_{13}$ are arbitrary. This leads to the computed pseudogap at $R_1$, $\Delta_{PG}(R_1)$, using Equation 1.

Figure 3(a) shows the degenerate $P_x$ and $P_y$ states that are occupied by two electrons in an isolated plaquette. Maintaining a uniform Fermi level throughout the crystal leads to the degenerate $P_x$ and $P_y$ levels in figure 3 at approximately the Fermi energy (see Supporting Information and S3 for details).

Figure 3(b) shows the coupling leading to the PG. The PG arises from the splitting of degenerate $P_x$ and $P_y$ symmetry Cu x2y2/O pσ states near the Fermi level that occur in an isolated plaquette. The splitting is due to the sum of the couplings of $P_x$ and $P_y$ states to each other that arise from second-order coupling through the other isolated plaquettes. For a given plaquette at location $R_1$, the magnitude of the second-order coupling through a different plaquette at position $R_2$ is,
$t_{xy}(R_1,R_2) = \Delta_0 \exp[-|R_1-R_2|/\xi] \exp[i\varphi(R_1,R_2)]$. The energy, $\Delta_0$, is on the order of the antiferromagnetic spin-spin coupling, $J_{dd} \approx$ 130 meV[40] and ξ is the spin correlation length seen by neutron scattering. Experimentally, ξ is found to approximately equal the mean distance between holes,[41] $\xi = a/x^{1/2}$, where *a* is the Cu-Cu distance in the $CuO_2$ planes. Finally, there is a random phase factor $\exp[i\varphi(R_1,R_2)]$ due to the random distribution of dopants (and hence plaquettes) in the crystal. The reason the spin-spin correlation length, ξ, appears in $t_{xy}(R_1,R_2)$ is because antiferromagnetic Cu $d^9$ spin regions exist between the plaquettes and the coupling of isolated plaquettes must occur through these regions. Thus, the coupling, $t_{xy}(R_1,R_2)$, attenuates with length scale ξ.

**Figure 4.** Calculated PG compared to experimental data of Tallon et al.[3] The data points are from $La_{2-x}Sr_xCuO_4$, $Y_{1-x}Ca_xBa_2Cu_3O_{7-\delta}$, and $Bi_2Sr_2CaCu_2O_{8+\delta}$ cuprates. The solid black line is calculated from Equation 1. The RMS error is ≈ 4 meV. The dashed curve is the superconducting gap, Δ, for $YBa_2Cu_3O_{7-\delta}$ where we use $2\Delta/kT_c = 3.5$ and obtain $T_c$ using the approximate equation,[42] $(T_c/T_{c,max})=1-82.6(x-0.16)^2$, where $T_{c,max} = 93$ K.

The PG for the plaquette at $R_1$ is given by the modulus of the sum, $t_{xy}(R_1) = \Sigma_{R_2} t_{xy}(R_1, R_2)$, $\Delta_{PG}(R_i)=<|t_{xy}(R_i)|>$. This can be approximated by the square root of the average of the modulus squared, $\Delta_{PG}(R_i) \approx <|t_{xy}(R_i)|^2>^{1/2}$, leading to (assuming uncorrelated phases)

$$\Delta_{PG}(R_i) = \Delta_0 \left( \sum_{j \neq i} e^{-2|R_i - R_j|/\xi} \right)^{\frac{1}{2}} \quad (1)$$

The average of Equation 1 over all isolated plaquettes leads to the PG seen in bulk measurements. For $\Delta_0 = 79.1$ meV, an excellent fit to experiment is obtained as seen in figures 4 and S3. The closing of the PG occurs at ≈ 0.19 because the distance between isolated plaquettes increases sharply in the vicinity of 0.19.

The Supporting Information contains a detailed discussion of the derivation of Equation 1 and also discusses the variation of the local PG with plaquette position.

In conclusion, we show that dopants in cuprates lead to the formation of four-site plaquettes localized in the vicinity of the dopants with a Cu/O $x^2-y^2$/pσ metallic band created in the percolating plaquette region. This quantitatively explains the doping evolution of the pseudogap as due to splitting of degenerate $P_x$ and $P_y$ states in isolated plaquettes.

ACKNOWLEDGEMENT We acknowledge discussions with Andrew D. Beyer, Andres Jaramillo-Botero, Eric W. Hudson, Carver A. Mead, and Slobodan Mitrovic. Funding was provided through gifts to the MSC.

SUPPORTING INFORMATION AVAILABLE Discussion of general doping of plaquettes, ab initio methods, out-of-plane hole characters, isolated plaquette orbitals, and analytic model for the pseudogap.

# Origin of the Pseudogap in High-Temperature Cuprate Superconductors

Jamil Tahir-Kheli and William A. Goddard III

*Materials and Process Simulation Center (MC 139-74)*
*California Institute of Technology, Pasadena CA 91125*

jamil@wag.caltech.edu, wag@wag.caltech.edu

**Supporting Information**

**Computational details**

Calculations to compute the pseudogap (PG) from Equation 1 were performed by averaging 100 ensembles of 2,500 x 2,500 lattices. Four-site plaquettes were randomly doped up to x = 0.187, then three-site plaquettes up to x = 0.226, followed by two-site plaquettes to x = 0.271, and finally one-site plaquettes were doped until no $d^9$ spins remained at x = 0.317. The details of the doping methodology are described in the manuscript.

The PG was determined for each isolated plaquette along with the average. The average value is the PG observed in bulk measurements (neutron, ARPES, specific heat, etc). STM sees an inhomogenous distribution of PG values. We find the standard deviation of the PG is approximately constant with doping and ~ 15 meV, in agreement with the STM value of ~ 10-20 meV.

Calculations to determine the doping values 0.187, 0.226, 0.271, and 0.317 at the boundaries of plaquette doping of four, three, two, and one undoped Cu $d^9$ spins were determined by generating a ensemble of 1,000 doped lattices of size 1,000 x 1,000 using a linear scaling percolation algorithm[1].

**Details of the ab-initio calculations that lead to out-of-plane hole character**

Pure density functionals such as local density (LDA)[2-4] and gradient-corrected functionals (GGA, PBE, etc)[5] obtain a metallic ground state for the undoped cuprates rather than an antiferromagnetic insulator. This is because pure density functionals underestimate the band gap due to a derivative discontinuity of the energy with respect to the number of electrons.[6,7] In essence, the LDA and PBE functionals include too much self-Coulomb repulsion. This leads to more delocalized electronic states in order to reduce this excess repulsion. Removing this extra repulsion is necessary to obtain the correct localized antiferromagnetic spin states of the undoped cuprates.

This particular problem with LDA has been known for a long time.[8] Very soon after the failure of LDA to obtain the undoped insulating antiferromagnet for cuprates, several approaches were applied to correct the flaws in LDA for $La_2CuO_4$. Using a self-interaction-corrected method[8] (SIC-LDA), Svane[9] achieved spin localization with an indirect band gap of 1.04 eV, and Temmerman, Szotek, and Winter[10] found a band gap of 2.1 eV. Using an LDA + U method, Czyzyk and Sawatzky[11] obtained 1.65 eV. In all of these calculations on the undoped cuprates, an increase in out-of-plane orbital character was noted in states just below the top of the valence band. Calculations with explicit dopants such as Sr in $La_{2-x}Sr_xCuO_4$ were not done.

Our ab-initio calculations[12,13] were performed using the hybrid density functional, B3LYP. B3LYP has been the workhorse density functional for molecular chemistry computations for almost 20 years due to its remarkable success on molecular systems.[14,15] For example,[14] B3LYP has a mean absolute deviation (MAD) of 0.13 eV, LDA MAD = 3.94 eV, and PBE MAD = 0.74 eV for the heats of formation, $\Delta H_f$, of the 148 molecules in the extended G2 set.[16,17] B3LYP has also been found to predict excellent band gaps for carbon nanotubes and binary and ternary semiconductors relevant to photovoltaics and thermoelectrics.[18,19]

The essential difference between B3LYP and LDA, PBE, and all pure density functionals is that 20% exact Hartree-Fock (HF) exchange is included. B3LYP is called a hybrid functional because it includes exact HF exchange. This removes some of the self-Coulomb repulsion of an electron with itself found in pure DFT functionals. A modern viewpoint of the reason for the success of hybrid functionals is that inclusion of some exact Hartree-Fock exchange compensates the error for fractional charges that occur in LDA, PBE, and other pure density functionals.[20] The downside to using hybrid functionals is they are computationally more expensive than pure density functionals.

Our B3LYP calculations reproduced the experimental 2.0 eV band gap for undoped $La_2CuO_4$ and also had very good agreement for the antiferromagnetic spin-spin coupling, $J_{dd}$ = 0.18 eV (experiment is ≈ 0.13 eV).[12] We also found substantial out-of-plane apical O and Cu z2 character near the top of the valence band in agreement with LDA + U and SIC-LDA calculations.

We also performed B3LYP calculations on $La_{2-x}Sr_xCuO_4$ for x = 0.125, 0.25, and 0.50 with explicit Sr atoms using large supercells.[13] Regardless of the doping value, we always found that the Sr dopant induces a localized hole in an out-of-the-plane orbital that is delocalized over the four-site region surrounding the Sr as shown in Figure 1 of the manuscript. This is in contrast to removing an electron from the planar Cu x2y2/O pσ as predicted by LDA and PBE.

Our calculations found that the apical O's in the doped $CuO_6$ octahedron are asymmetric anti-Jahn-Teller distorted. In particular, the O atom between the Cu and Sr is displaced 0.24 Å while the O atom between the Cu and La is displaced 0.10 Å. XAFS measurements[21] find the apical O displacement in the vicinity of a Sr to be ≈ 0.2 Å.

**Description of the out-of-plane plaquette orbital obtained from ab-initio QM calculations**

Figure 1 of the main text shows the out-of-plane localized hole orbital we obtained in the vicinity of a dopant atom.[12,13,22-24] In $La_{2-x}Sr_xCuO_4$, the formal valence of Sr is +2 and La is +3. This means the environment in the vicinity of Sr is more negatively charged than the average La environment in the crystal. Removing an electron from an out-of-the-plane orbital in the vicinity of the Sr reduces this excess Coulomb repulsion.

Since our calculations had a fixed Néel antiferromagnetic background for the undoped Cu $d^9$ sites, we found two degenerate states with the plaquette localized as shown in Figures 1(a) and (b). The spin of the plaquette hole state was opposite in the $P_{x'}$ and $P_{y'}$ states due to the fixed Neel antiferromagnetic background (not shown in figure). Exchange coupling of the electron spin of the occupied plaquette state, $P_{x'}$ or $P_{y'}$, makes its spin equal to the spin of the singly occupied x2y2 spin at the Cu site below the apical O $p_z$.

In our prior work, we assumed that computations with a more realistic background of antiferromagnetic spins would lead to delocalization of the plaquette out-of-plane orbital over all four Cu sites in the plaquette.[12,13,22-24] Thus, including spin, there are four possible out-of-plane plaquette states that are delocalized over a 4-site Cu square. These four states are occupied by three electrons.

The two possible plaquette hole orbital states are shown in figures 1(b) and (c). They have $P_{x'}$ and $P_{y'}$ symmetry and are degenerate in energy if there is only a single dopant in the material and there is no long range antiferromagnetic order in the undoped $d^9$ region as is found for the superconducting range of dopings. Due to the random dopant environment, the $P_{x'}$ and $P_{y'}$ states can mix to form two orthogonal states with a small energy splitting. The random distribution of these energy splittings in the system is the source of the linear resistivity in our model.[23]

The $P_{x'}$ and $P_{y'}$ out-of-plane plaquettes states above and shown in Figure 1 are different from the delocalized (inside an isolated plaquette) Cu x2y2/O pσ states with the same symmetry, $P_x$ and $P_y$ shown in Figures 3 and S2 (the x2y2/pσ states can be transformed to $P_{x'}$ and $P_{y'}$). It is the latter states that split and lead to the PG. The reason the out-of-plane $P_{x'}$ and $P_{y'}$ energy splitting is not the PG is because they are hole states at ≈ 0.1 eV above the Fermi level as seen in the density of states figures of Perry et al.[13]

**Formation of the standard planar Cu x2y2/O pσ band inside the 3D percolating plaquette swath**

The Cu x2y2/O pσ states inside 4-site plaquettes are expected to delocalize. The reason for this delocalization is described below.

In models where the O sites are renormalized out, there is a Hubbard on-site Coulomb repulsion, U, and a Cu-Cu hopping matrix element, $t_{eff}$. In these models, $t_{eff} \ll U$, leading to localization of spins on the Cu sites.

When the realistic picture that includes the O sites is used, the effective hopping matrix element, $t_{eff}$, is given by,

$$t_{eff} \sim \frac{t^2}{(\varepsilon_d + U - \varepsilon_p)}$$

where t is the Cu-O hopping and $\varepsilon_d$ and $\varepsilon_p$ are the Cu x2y2 and O pσ orbitals energies, respectively. This scenario is shown on the left-side of Figure 1. Localization occurs when $t_{eff} \ll \varepsilon_d + U - \varepsilon_p$.

The affect of creating a hole in an out-of-plane orbital, as shown in Figure 1, is to lower the orbital energies $\varepsilon_d$ and $\varepsilon_p$. Since the largest hole character is on the apical O $p_z$ that resides directly above the Cu atoms and the Cu dz2, the Cu x2y2 orbital energy, $\varepsilon_d$, is lowered much more than the O pσ orbital energy, $\varepsilon_p$. This brings the upper Hubbard Cu orbital energy, $\varepsilon_d + U$, closer to the O orbital, $\varepsilon_p$. The effective hopping matrix element increases and the $d^9$ spins on the doped Cu sites delocalize. When the plaquettes percolate in 3D through the crystal, the standard Cu x2y2/O pσ band is formed on these doped "four-site" regions (the percolating plaquette swath shown in Figures 2 and S1).

Our explanation for delocalization depends on the detailed differences between the planar Cu and O orbital energies. Renormalizing away the O sites prior to considering the affect of the out-of-plane hole leads to no delocalization because there is no change in the effective hopping, $t_{eff}$.

Cu x2y2/O pσ delocalization in the plaquettes interacting with localized Cu $d^9$ spin in the undoped regions leads to explanations for the doping phase diagram, the neutron resonance peak with doping, the dispersionless incommensurabilities found in STM, and the universal room-temperature theromopower[23,24] using simple counting arguments. The first three properties were obtained with no adjustable parameters while the thermopower required exactly one parameter. A large spectrum of additional cuprate phenomenology[22,23] was also qualitatively explained.

**Discussion of random placements of plaquettes as a function of doping**

Since the dopants minimize their Coulomb repulsion with each other (screened by the percolating metallic electrons), we assume dopants are distributed in the crystal with the constraint of no plaquette overlaps, but otherwise the distribution is completely random. Above ≈ 0.187 doping, it is impossible to add dopants such that the corresponding 4-site plaquette does not overlap (does not share a corner Cu) with a another plaquette. Given that plaquette overlap can no longer be avoided above x ≈ 0.187, the most energetically favorable location for a dopant to reside is on a site that leads to a 4-site plaquette doping three Cu $d^9$ atoms and one doped Cu atom. This is shown as green squares in Figures 2 and S1. At x ≈ 0.226 doping, it is no longer possible to locate three undoped Cu $d^9$ sites that are part of a 4-site square. Hence, the most energetically favorable sites are those that dope two Cu $d^9$ and two doped Cu. These plaquettes are shown in pink in Figures 2 and S1. This form of doping can continue to x ≈ 0.271. At this point, there are no adjacent pairs of $d^9$ spin remaining. Single Cu $d^9$ site doping can continue out to ≈ 0.317. Above this value, there are no remaining localized Cu $d^9$ spins remaining in the material. The percolating metallic region has become the whole crystal. Figure S1 depicts the undoped $d^9$ sites and plaquettes as a function of doping for 0.05 < x < 0.32 and S4 plots the doping evolution of the number of metallic and isolated 4-site plaquettes. At dopings x ≈ 0.05−0.06, we calculate that the plaquettes percolate in 3D,[1] leading to the formation of the standard Cu x2y2/O pσ metallic band inside the percolating "metallic" swath. Since superconducting pairing occurs at the surface where there are adjacent $d^9$ spins (green and blue arrows in Figures 2 and S1) and not for isolated spins (pink arrows), superconductivity vanishes at x ≈ 0.271 in agreement with the experimental value of ≈ 0.27.

**Evolution of isolated 4-site plaquettes with doping**

**Figure S1** (shown on pages 9-11) Schematic of a single $CuO_2$ plane for twelve different doping values on a 20 x 20 lattice. The dopings span from x = 0.05 to x = 0.32. The black dots are undoped Cu $d^9$ spins that antiferromagnetically couple to each other. The squares are 4-site plaquettes with Cu atoms at the corners. The out-of-plane orbital and the O atoms are not shown. The black squares are plaquettes that are adjacent (along the Cu-O bond directions) to another plaquette. The red squares are isolated 4-site plaquettes. A delocalized metallic Cu x2y2/O pσ band is formed on the percolating swath of plaquettes for dopings larger than ≈ 0.05. This is shown here in yellow. For dopings less than ≈ 0.151, the percolation occurs in 3D and cannot be seen in this figure. It occurs through coupling of yellow regions to $CuO_2$ layers above and below the $CuO_2$ layer that is shown here. Above x = 0.151, 2D percolation occurs and this can be seen directly in the figures. Above 0.187 doping, additional plaquettes must overlap another plaquette. The green squares are 4-site plaquettes that dope three Cu $d^9$ spins. This occurs from x ≈ 0.187 to x ≈ 0.226. Above 0.226, only two Cu $d^9$ spins can be doped for each new plaquette. They are shown by pink squares. Above 0.271 doping, only single Cu $d^9$ spins exist and the blue squares represent the plaquette doping that includes one Cu $d^9$ spin. There exist undoped adjacent $d^9$ pairs in the x = 0.26 figure, but in the x = 0.27 figure there are no remaining antiferromagnetic $d^9$ adjacent pairs to cause superconducting pairing. Thus, the superconducting phase ends at x ≈ 0.271 (experiment is ≈ 0.27). Plaquette doping of a single Cu $d^9$ (blue squares) can occur up to ≈ 0.317 doping. The final figure at x = 0.32 shows a fully doped crystal with no remaining localized Cu $d^9$ spins. At this doping, all planar atoms are in the metallic swath. Further doping does not increase the metallic region.

The isolated 4-site plaquettes (red squares) decrease as the doping increases. The density of isolated plaquettes declines sharply at x ≈ 0.19. By x = 0.21 there are no isolated plaquettes in the 20 x 20 lattice here. Isolated plaquettes will always exist in an infinite crystal, but the distance between them becomes so large that they are much farther apart than the neutron spin correlation length, $\xi = a/\sqrt{x}$. This leads to the PG going to zero at x ≈ 0.19 as is experimentally observed. This doping value is independent of the magnitude of the isolated plaquette to isolated plaquette coupling ($\Delta_0$ in manuscript). It is a purely geometric consequence of the counting of 4-site plaquettes and their doping.

**x = 0.05** 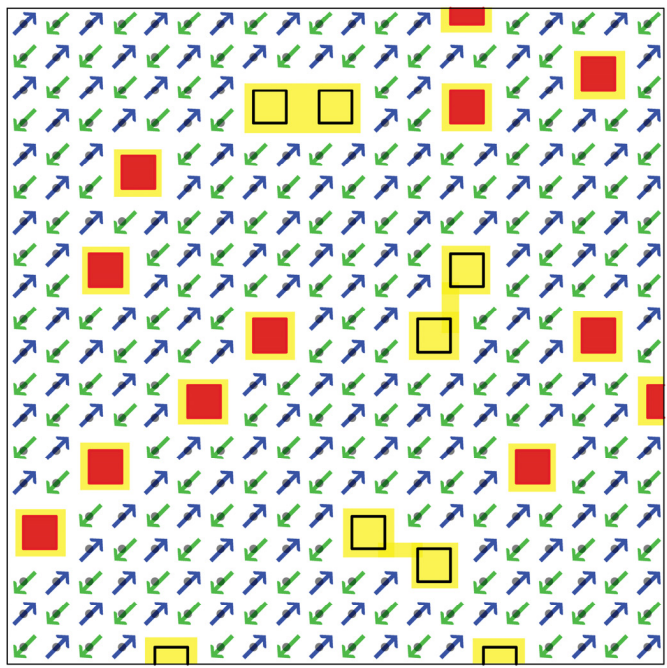

**x = 0.075** 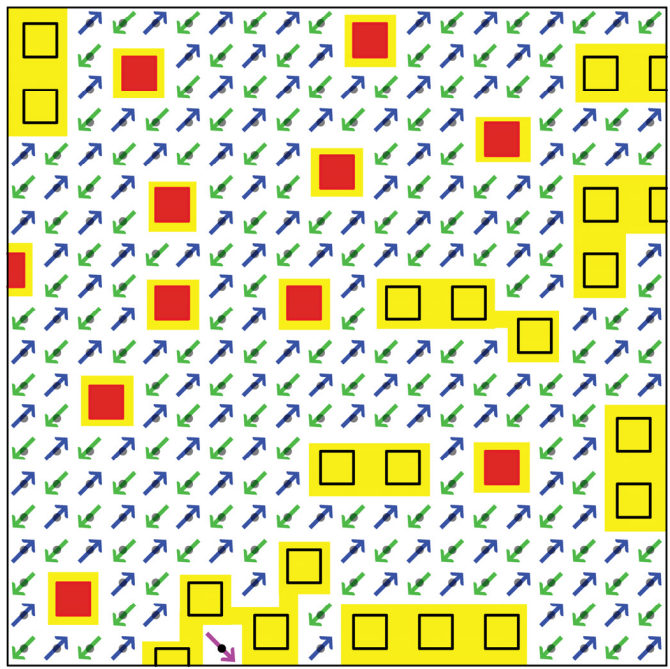

**x = 0.10** 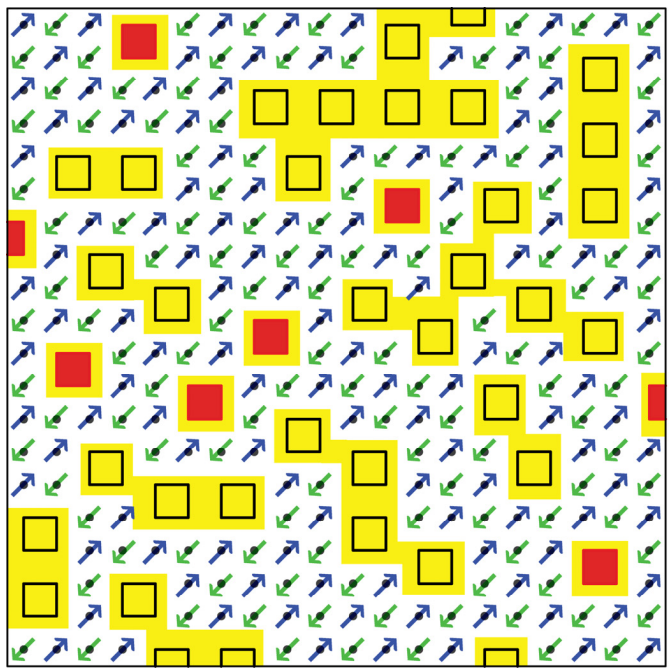

**x = 0.125** 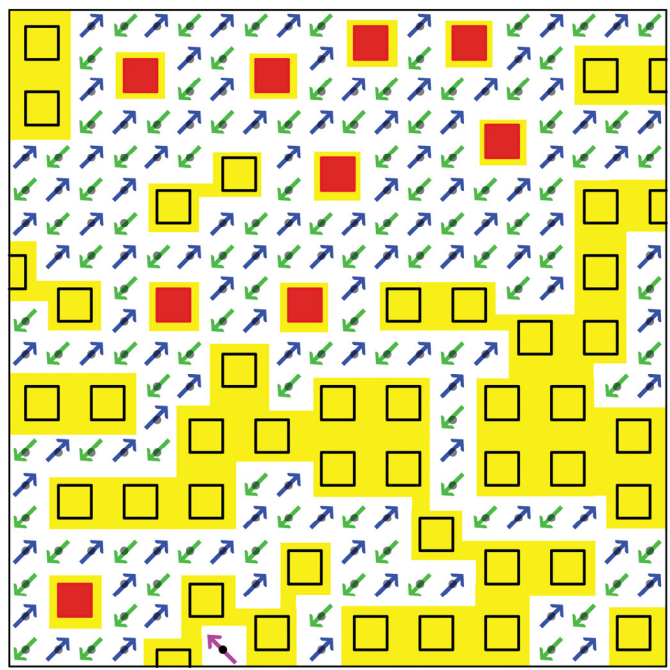

The Figure caption is on page 7

| **x = 0.15** | **x = 0.18** |
|---|---|
| 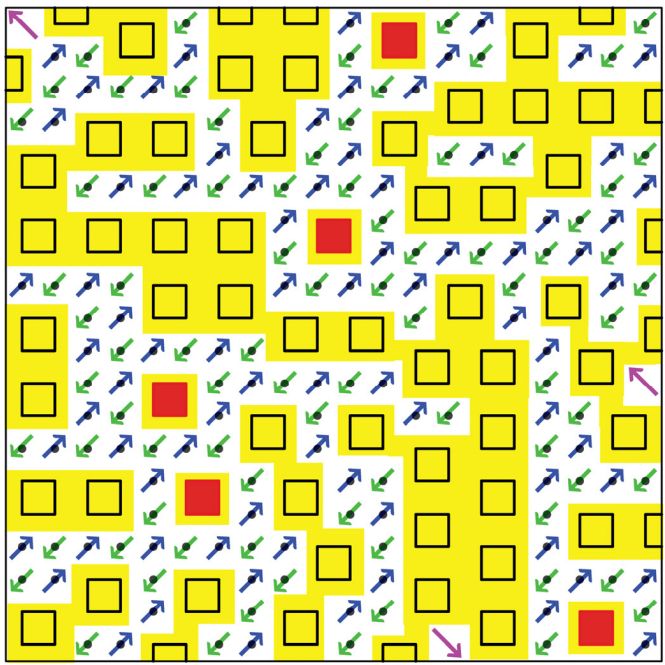 | 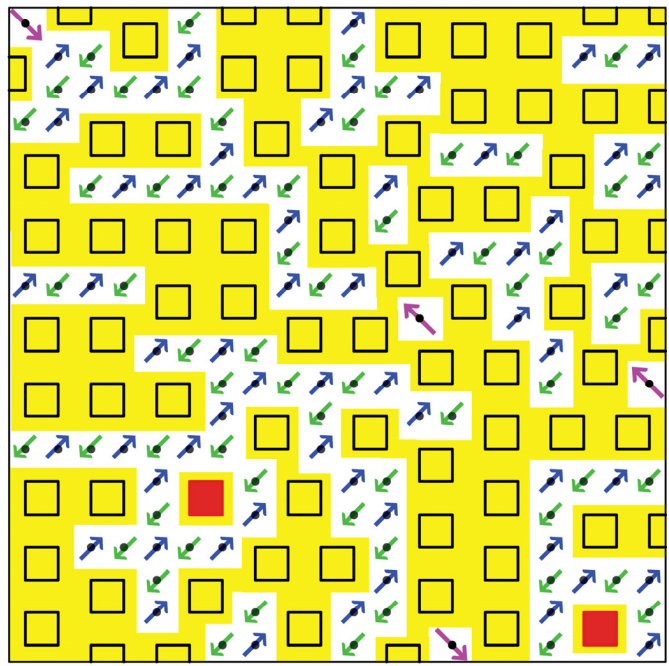 |
| **x = 0.19** | **x = 0.20** |
| 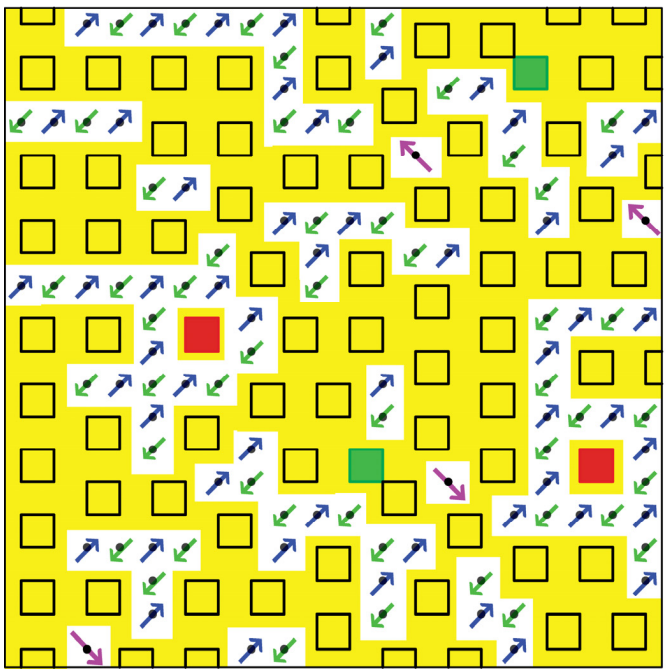 | 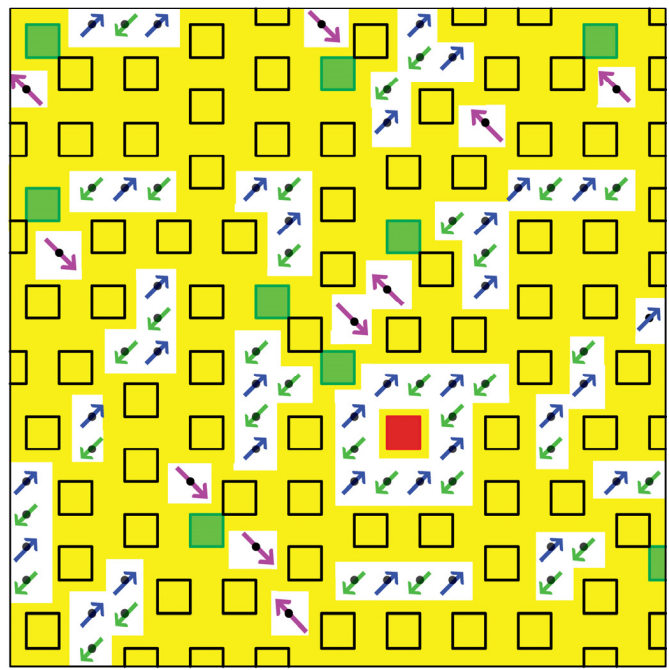 |

The Figure caption is on page 7

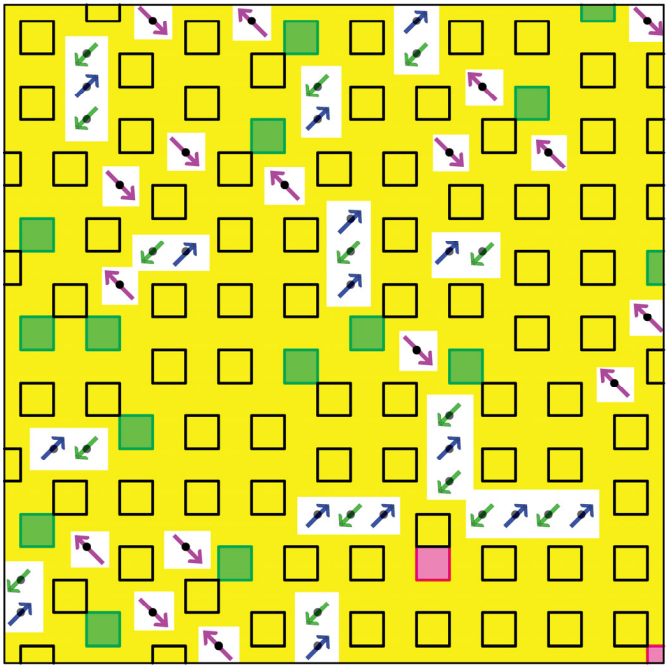
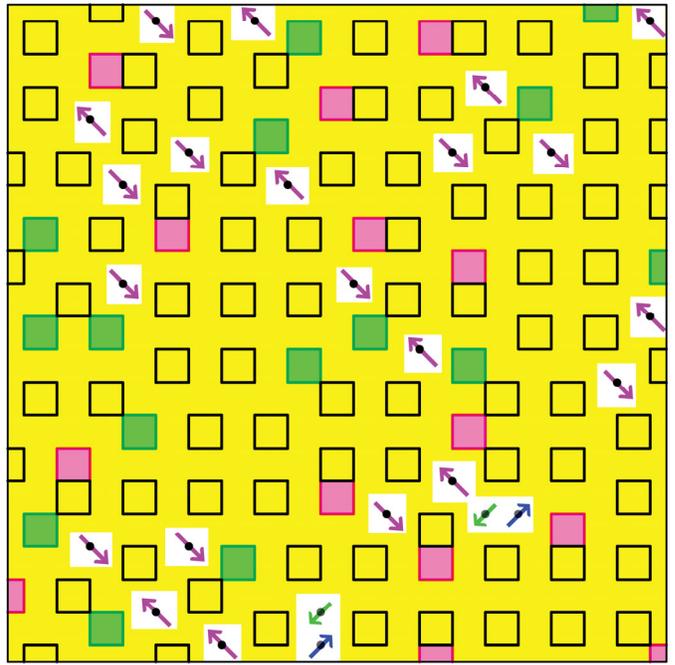
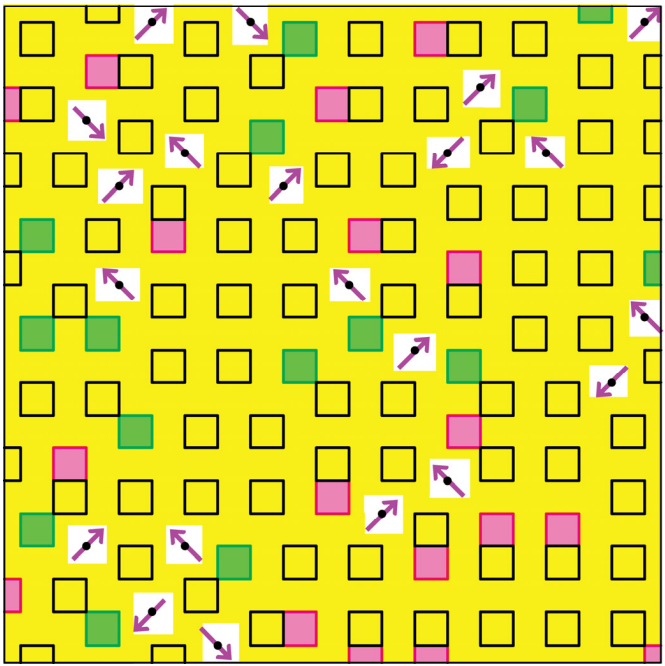
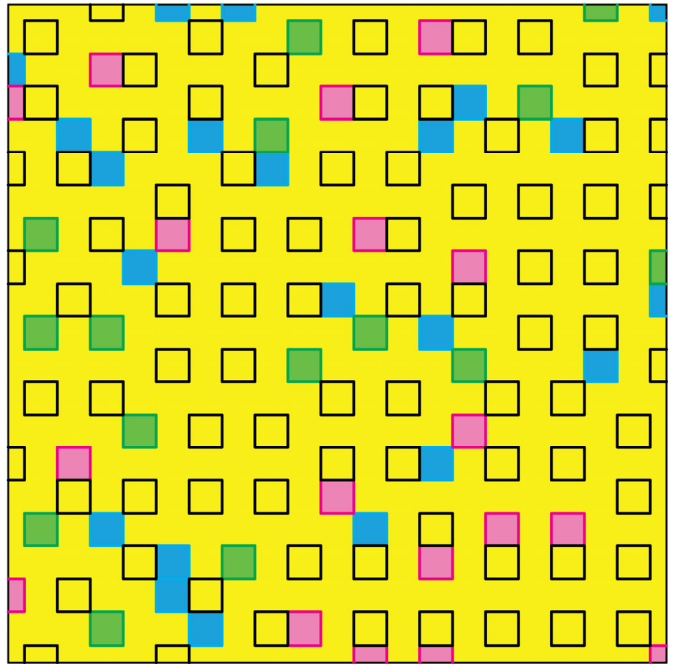

The Figure caption is on page 7

**Approximate pinning of $P_x$ and $P_y$ states near Fermi level**

The eight possible delocalized Cu x2y2/O pσ eigenstates inside an isolated 4-site plaquette are shown in Figure S2. The states are arranged with the highest energy state at the top and the lowest energy state at the bottom. There are two electrons per O pσ and one electron per Cu x2y2 leading to twelve total electrons. This leaves the degenerate $P_x$ and $P_y$ anti-bonding states with two electrons. Expanding the $P_x$ and $P_y$ states as band k-states, we find $P_x$ has the largest overlap with k = (0, π) and $P_y$ with k = (π, 0). These k states are always found to be close to the Fermi level in cuprates. The energies of these isolated plaquettes states will be shifted by the electrostatic potential inside the crystal arising from the rearrangement of charges that occur at equilibrium when the Fermi energy must become constant throughout the crystal. Since the antiferromagnetic $d^9$ region has a charge density of one electron per Cu and two electrons per O, the charge density inside isolated plaquettes will be approximately the same. This leads to the Fermi level being placed very close to the degenerate anti-bonding $P_x$ and $P_y$ states. The Fermi level is approximately "pinned" near the degeneracy.

Since the anti-bonding Px and Py states have predominantly (π, 0) and (0, π) character, the PG will appear in angle-resolved-photoemission (ARPES) to be largest at these k-vectors.

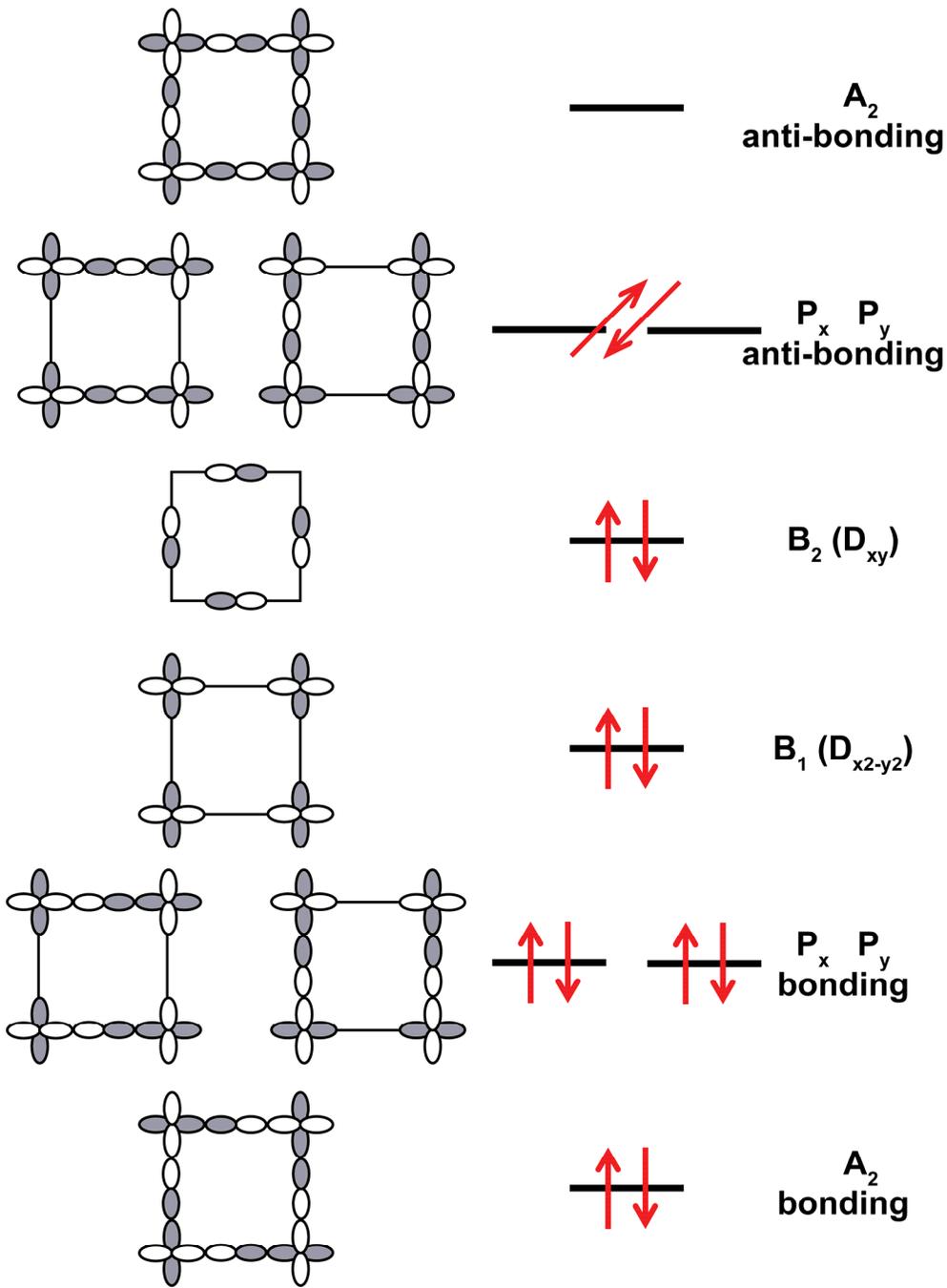

**Figure S2** The eight possible delocalized Cu x2y2/O pσ eigenstates inside an isolated 4-site plaquette. The states are arranged with the highest energy state at the top and the lowest energy state at the bottom. There are two electrons per O pσ and one electron per Cu x2y2 leading to twelve total electrons. There are only two electrons to occupy the four degenerate $P_x$ and $P_y$ anti-bonding states. Theses states have dominant k-vector character at $(\pi, 0)$ and $(0, \pi)$, respectively. The splitting of the $P_x$ and $P_y$ states leads to the PG.

**Discussion of the diagonal and off-diagonal terms in the 2 x 2 Hamilton matrix for the PG of a particular isolated 4-site plaquette**

For a given isolated plaquette located at position **R**, the Hamiltonian matrix for the $P_x$ and $P_y$ orbitals is $H_0 \approx \begin{bmatrix} \varepsilon_F & 0 \\ 0 & \varepsilon_F \end{bmatrix}$, where $\varepsilon_F$ is the Fermi energy. The perturbing matrix for coupling of this plaquette to the antiferromagnetic $d^9$ spins, the delocalized Cu x2y2/O pσ swath, and other isolated plaquettes is $H' \approx \begin{bmatrix} \varepsilon_{xx} & t_{xy}(R) \\ t_{xy}(R) & \varepsilon_{yy} \end{bmatrix}$. We expect the diagonal terms to be approximately the same, $\varepsilon_{xx} \approx \varepsilon_{yy}$, due to the random doping environment. Any non-zero value leads to a particle-hole asymmetry of the PG as is observed.[25,26] Thus the modulus of the off-diagonal term, $|t_{xy}|$, determines the size of the splitting and the PG for this isolated plaquette.

The first expectation is that the dominant coupling of the $P_x$ state to $P_y$ is by second-order coupling through the antiferromagnetic spins neighboring the plaquette. Since the plaquette is isolated, it is completely surrounded by antiferromagnetic spins in the plane and the approximate $D_4$ symmetry of the plaquette is maintained by including these antiferromagnetic spins. Thus coupling to the antiferromagnetic spins cannot split the degeneracy of the $P_x$ and $P_y$ states.

Coupling of $P_x$ and $P_y$ can occur by interaction through the antiferromagnetic $d^9$ spins to the delocalized metallic Cu x2y2/O pσ states in the percolating plaquette swath. Since the bandwidth of the metallic band inside the percolating swath is ≈ 2.0 eV and the shape of the metallic swath is random, this coupling should be small.

The largest coupling between $P_x$ and $P_y$ inside a particular isolated plaquette is with states close in energy to $P_x$ and $P_y$ (since the energy difference appears in the denominator in a second-order perturbation). This leads to the dominant coupling through other isolated plaquettes. For two isolated plaquettes to interact, they must couple through the antiferromagnetic $d^9$ spins. Due to the finite correlation length of the antiferromagnetic spins, ξ, this coupling will be exponentially attenuated for distances larger than $\xi$. Thus the magnitude of the PG is a function of the distances between isolated clusters as a function of doping as shown in Figure 3b.

The off-diagonal coupling matrix element, $t_{xy}(\mathbf{R_i})$, where $\mathbf{R_i}$ is the position of the $i^{th}$ isolated plaquette, is the sum of terms, $t_{xy}(\mathbf{R_i}, \mathbf{R_j})$, where $\mathbf{R_j}$ is the position of the $j^{th}$ isolated plaquette

$$t_{xy}(R_i, R_j) = \Delta_0 e^{-|R_i - R_j|/\xi} e^{i\varphi(R_i, R_j)}$$

$$t_{xy}(R_i) = \sum_j t_{xy}(R_i, R_j)$$

The first term in $t_{xy}(\mathbf{R_i}, \mathbf{R_j})$ is the constant, $\Delta_0$. It sets the energy scale of the coupling and should be on the order of the antiferromagnetic spin-spin coupling, $J_{dd} = 130$ meV. The first exponential is the damping of the matrix element due to transmission through the $d^9$ antiferromagnetic region with finite correlation length, $\xi$. The ratio of the distance between the plaquettes, $|\mathbf{R_i} - \mathbf{R_j}|$, and the correlation length determines the magnitude of the damping. The final term is the phase.

Due to the arbitrary distribution of the plaquettes and the antiferromagnetic $d^9$ environment, we take the phases coupling isolated plaquettes at $\mathbf{R_j}$ and $\mathbf{R_k}$ to the isolated plaquette at $\mathbf{R_i}$ to be uncorrelated, $\langle \varphi(R_i, R_j) \varphi(R_i, R_k) \rangle = 0$, for $j \neq k$.

The PG is given by one-half of the splitting of the $P_x$ and $P_y$ levels, $\Delta_{PG}(R_i) = |t_{xy}(R_i)|$. Due to the random phases, $\varphi(R_i, R_j)$, the PG is the average, $\Delta_{PG}(R_i) = \langle |t_{xy}(R_i)| \rangle$. The average of the modulus can be well approximated by the square root of the average of the modulus squared, leading to

$$\Delta_{PG}(R_i) = \sqrt{\langle |t_{xy}(R_i)|^2 \rangle}$$

$$\Delta_{PG}(R_i) = \Delta_0 \left( \sum_j e^{-2|R_i - R_j|/\xi} \right)^{\frac{1}{2}} \quad (1)$$

Neutron spin scattering experiments find the correlation length to be approximately equal to the mean spacing of holes, $\xi \approx a/\sqrt{x}$, where $a$ is the nearest-neighbor Cu-Cu distance (a ≈ 3.8 Å) and $x$ is the doping.[27] The above expression is used to compute the PG curve shown in the main text.

**An analytic expression for the PG**

An approximate analytic expression for $\Delta_{PG}$ can be obtained from the computed number of isolated plaquettes as a function of doping, $N_4$. Due to the exponential decay of the matrix element with length scale, $\xi$, only isolated plaquettes inside an area on the order of $\sim \xi^2$ can contribute to the sum in Equation 1. Let $a$ be the Cu-Cu distance in the plane and N equal the total number of Cu sites in the plane. Then the average number of isolated plaquettes in the area A is $A\left(\dfrac{N_4}{Na^2}\right)$. Substituting $A = \xi^2$ leads to the approximate analytic expression

$$\Delta_{PG} = \Delta_0' \sqrt{\dfrac{N_4 \xi^2}{Na^2}} = \Delta_0' \sqrt{\dfrac{N_4}{Nx}}$$

$\Delta_0'$ is a constant close to $\Delta_0$. Figure S3 shows the fit to the analytic expression above. The fit is quite good, but the exact evaluation of Equation 1 leads to a better fit.

**Inhomogeneous PG determination**

Equation 1 leads to a different value for the PG for each isolated plaquettes. STM measurements observed an inhomogeneous distribution of the PG with an approximately doping independent standard deviation of ~ 10 – 20 meV. Our calculations also find an approximately constant value of ~ 15 meV.

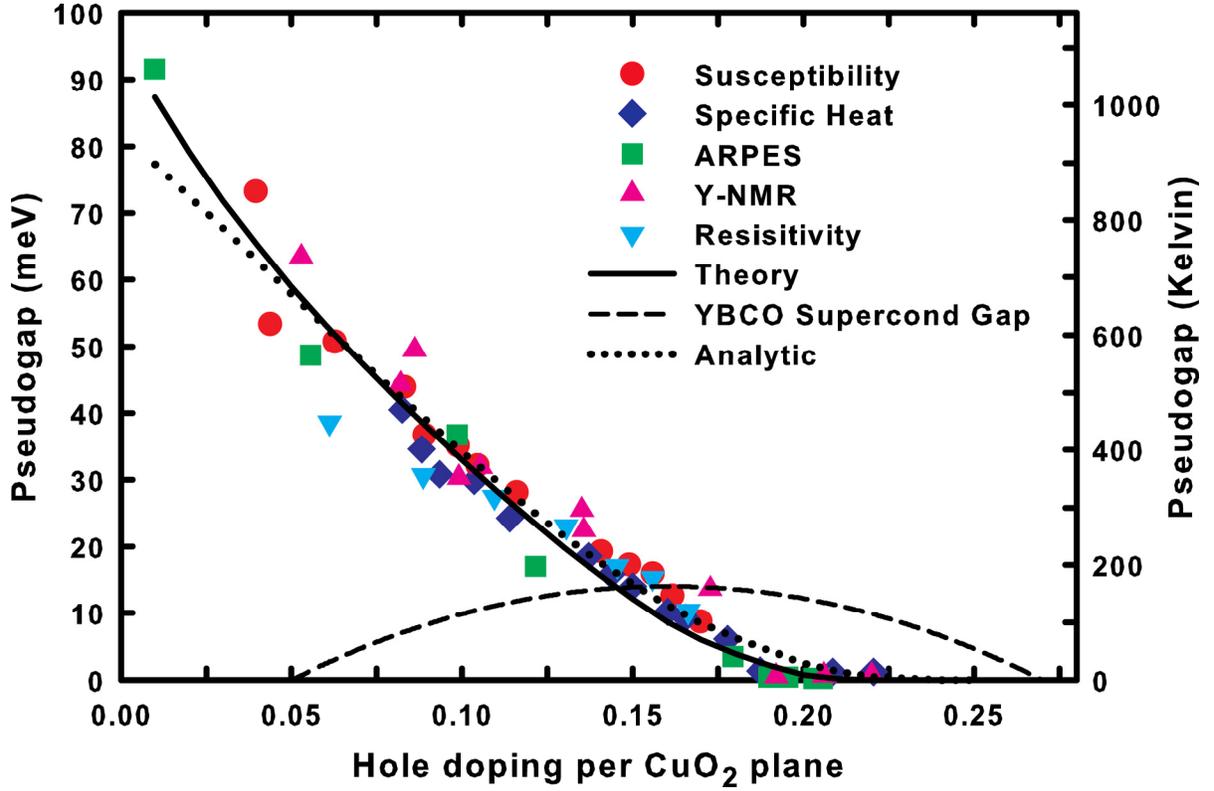

**Figure S3.** Comparison of calculated PG curve using Equation 1 and the approximate analytic expression, $\Delta_{PG} = \Delta'_0 \sqrt{N_4/Nx}$, where $\Delta'_0 = 82.4$ meV. Here, $N_4$ is the total number of isolated plaquettes at doping, x. N is the total number of Cu atoms. The dotted curve is the analytic expression and the solid black curve is calculated from Equation 1. The dashed curve is the superconducting gap, $\Delta$, for $YBa_2Cu_3O_{7-\delta}$ where we use $2\Delta/kT_c = 3.5$ and obtain $T_c$ using the approximate equation,[28] $(T_c/T_{c,max}) = 1 - 82.6(x - 0.16)^2$, where $T_{c,max} = 93$ K. The PG from the analytic expression remains non-zero for slightly higher dopings than 0.20 and does not obtain as large of a PG for low doping as the computed result from Equation 1. The data points in the figure are taken from Tallon et al.[29]

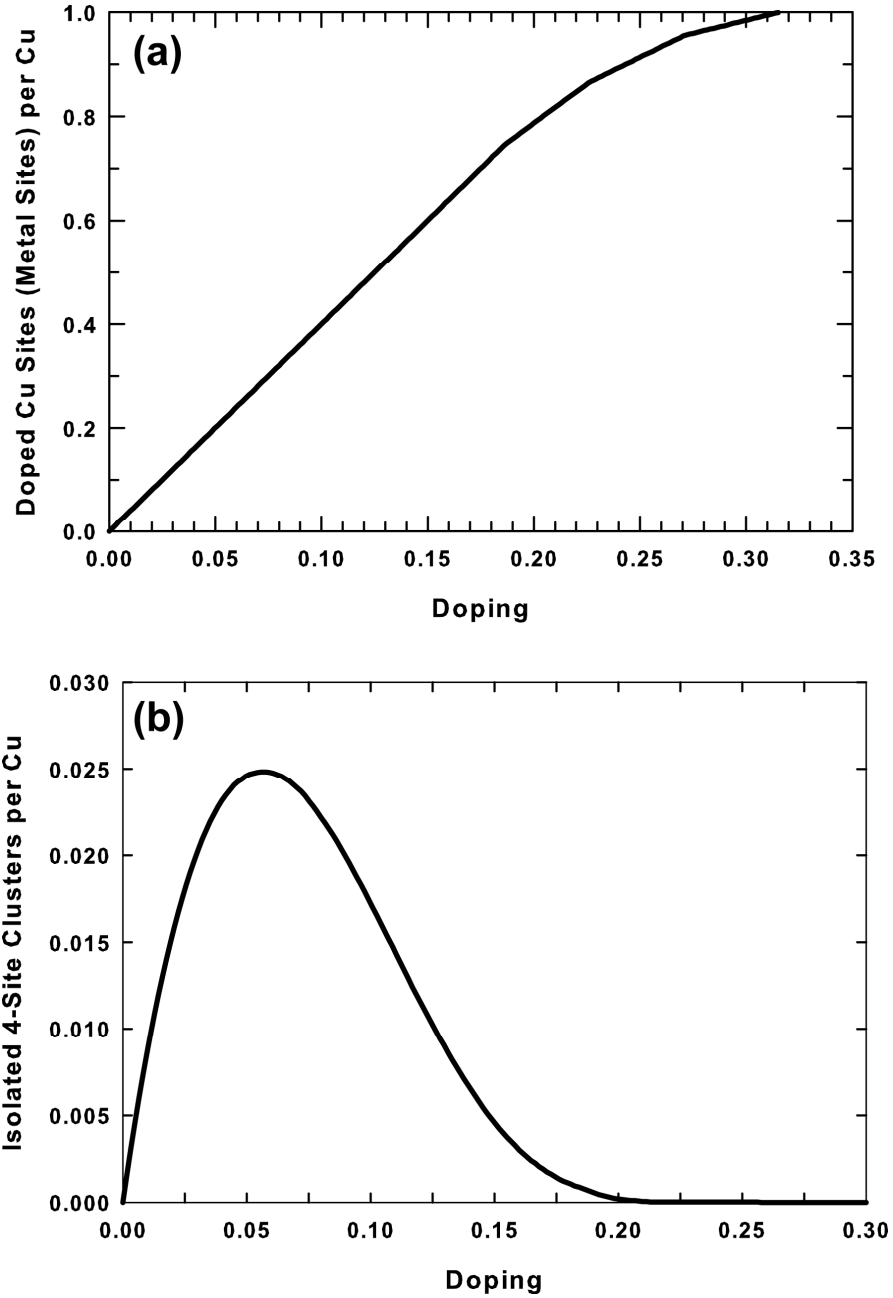

**Figure S4.** The evolution of the metal sites and isolated plaquettes with doping. In (a), the number of doped Cu sites per total Cu sites is plotted as a function of doping. Above ≈ 0.05 doping, when the plaquettes percolate through the crystal, a metallic Cu x2y2/O pσ band is formed in the percolating swath. The curve increases as $4x$ up to ≈ 0.187 because each plaquette dopes four Cu sites. Further doping increases the doped Cu sites by $3x$ up to ≈ 0.226, by $2x$ from 0.226 – 0.271, and then by $x$ until 0.317 when all Cu sites become doped (metallic). (b) shows the number of isolated 4-site clusters as a function of doping. For extremely low doping, all plaquettes are isolated and the curve increases as $x$.

There is a peak at ≈ 0.058. At x ≈ 0.20, the number of isolated plaquettes becomes almost zero and the PG vanishes.